\documentclass[12pt]{article}
\usepackage[margin=0.95 in]{geometry}
\usepackage{amsmath} 
\usepackage{amssymb,amsfonts}
\usepackage[all]{xy}
\usepackage{graphicx}
\usepackage[utf8x]{inputenc}
\usepackage{amsmath}
\usepackage{amssymb}
\usepackage{float}
\usepackage{array}
\usepackage{tikz}
\usepackage{mathtools}
\usepackage{mathrsfs} 
\usepackage[hidelinks]{hyperref}
\usepackage{cite}
\numberwithin{equation}{section}
\numberwithin{equation}{section}
\numberwithin{table}{section}
\begin{document}

\thispagestyle{empty}

\vspace*{3cm}
{}

\noindent
{\LARGE \bf Twisted  String Theory in Anti-de Sitter Space}
\vskip .4cm
\noindent
\linethickness{.06cm}
\line(10,0){460}
\vskip 1.1cm
\noindent
\noindent
{\large \bf Songyuan Li and Jan Troost}
\vskip 0.25cm
{\em 
\noindent
 Laboratoire de Physique de l’\'Ecole Normale Sup\'erieure \\ 
 \hskip -.05cm
 CNRS, ENS, Universit\'e PSL,  Sorbonne Universit\'e, Universit\'e de Paris \\
 \hskip -.05cm
 F-75005 Paris, France
}
\vskip 1.2cm

\vskip0cm

\noindent
{\sc Abstract: } {We construct a string theory in three-dimensional anti-de Sitter space-time that is independent of the boundary metric. It is a topologically twisted theory of quantum gravity. We study  string theories  with an asymptotic $N=2$ superconformal symmetry and demonstrate that, when the world sheet coupling to the space-time boundary metric undergoes a $U(1)$ R-symmetry twist,  the space-time boundary energy-momentum tensor becomes topological. As a by-product of our analysis, we obtain the world sheet vertex operator that codes the space-time energy-momentum  for  conformally flat boundary metrics. }

\vskip 1cm

\pagebreak

\newpage
\setcounter{tocdepth}{2}
\tableofcontents

\section{Introduction}
\label{introduction}
String theory in anti-de Sitter space-time is conjectured to be holographically dual
to conformal field theory \cite{Maldacena:1997re}. The most prominent examples of the conjecture have more than one supersymmetry.
The corresponding boundary conformal field theories allow for a topological twist \cite{Witten:1988ze} and therefore, by holographic duality, so does string theory in anti-de Sitter space-time. Thus, topologically twisting quantum theories of gravity in anti-de Sitter space-time is a natural enterprise.

At the supergravity level, there have been multiple contributions to understanding the twisted theory in the bulk. Bulk supergravity path integrals have been shown to be amenable to localization. See  e.g. \cite{Dabholkar:2014ema,Bonetti:2016nma,BenettiGenolini:2017zmu,BenettiGenolini:2018iuy,deWit:2018dix,BenettiGenolini:2019jdz}. The approach to the twisting of supergravity in \cite{Costello:2016mgj} on the other hand, is based on a BCOV type topological theory \cite{Bershadsky:1993cx} generalized to arbitrary Calabi-Yau manifolds \cite{Costello:2012cy}.\footnote{The topological string theory in question has been constructed in a mathematical framework. It would be interesting to have a physical  understanding of the theory close  to the  BCOV derivation of topological string theory on Calabi-Yau three-folds.}  It has been used to study the backreaction of D-branes in a topological string theory in the spirit of the original derivation of the AdS/CFT correspondence
\cite{Costello:2018zrm}, including in three-dimensional anti-de Sitter space-time \cite{Costello:2020jbh}. Moreover, string theory on $AdS_3$ at a curvature radius equal to the string length has  been demonstrated to have topological features, and has been holographically matched to an impressive degree \cite{Eberhardt:2019ywk,Eberhardt:2020akk}. Other approaches to  topologically twist $AdS_3$ string theory include   \cite{Sugawara:1999fq,Rastelli:2005ph}.

In this paper, we follow the conceptual road laid out in \cite{Li:2019qzx}, where we exploited a holographic duality between   Chern-Simons supergravity in the bulk and a boundary conformal field theory with extended supersymmetry. In this simplified model of quantum  holography, we showed how to obtain a topologically twisted bulk theory. It was constructed by noting that the topological boundary energy-momentum tensor couples in a twisted manner to the boundary degrees of freedom, and that therefore the introduction of a non-trivial boundary metric  requires twisted asymptotic boundary conditions in the bulk. These twisted asymptotic boundary conditions, combined with a suitably defined cohomology of physical states, define a bulk theory independent of the boundary metric. In that sense, it is a topologically twisted theory of quantum gravity \cite{Li:2019qzx}.

In this paper, we define a string theory in three-dimensional anti-de Sitter space-time with  specific couplings to the boundary metric. The bulk generating function of correlation functions is defined as a functional of the boundary metric.  The derivative of the generating function with respect to the boundary metric is the boundary energy-momentum tensor. Our goal will be to demonstrate that with an appropriate twist in the boundary conditions, the boundary energy-momentum tensor is the energy-momentum tensor of the topologically twisted boundary theory. Since the latter is known to be cohomologically exact, that guarantees the independence of the bulk correlators from the boundary metric.
Thus, we take an important  step in understanding the bulk topological string theory.

Our paper is structured as follows. In section \ref{Generating} we review the string generating function of correlation functions. We emphasize that the string background around which we expand must be on-shell, and that the choice of background solution, and in particular its boundary conditions, determines the resulting boundary energy-momentum tensor.
In section \ref{Solutions} we describe three-dimensional anti-de Sitter string theory backgrounds with purely NS-NS flux for a generic boundary metric. Next, we generalize these string backgrounds to include an extra circle. The circle is a geometric counterpart to the $U(1)_R$ symmetry that we need to topologically twist the boundary superconformal field theory. We render the dependence of the $AdS_3$ string background on the boundary metric  explicit. We also present the bulk solution with a boundary  gauge field that couples to the R-symmetry current.
In section \ref{Tensors}, we compute the world sheet vertex operators that correspond to the physical boundary energy-momentum tensor, the R-symmetry current, as well as the topological boundary energy-momentum tensor. The expressions are derived for a generic conformally flat boundary.  Finally, we  prove in a simple manner that the boundary energy-momentum tensor for the string theory with appropriately twisted boundary conditions is indeed topological.
We conclude with a summary and suggestions for future research in section \ref{Conclusions}.

\section{The Generating Function}
\label{Generating}
\label{GeneratingFunction}
We define the string theory generating function
$Z$ of correlation functions:
\begin{equation}
Z[\phi^{(0)},J] = \sum_{{\text{genus}}=0}^\infty
Z_{\text{genus}}[\phi^{(0)},J] = \sum_{{\text{genus}}=0}^\infty \int [DX]_{\text{genus}} \exp (- S_{}[\phi^{(0)},J,X]) \, .     \label{DefinitionGeneratingFunction}
\end{equation}
The  generating function $Z$ has a perturbative definition as a sum over genera. 
We emphasized that the generating function of correlation functions depends on the background  fields that we collectively denoted $\phi^{(0)}$. The generating function is also a functional of the sources $J$ which couple to  the normalizable string excitations. The correlation functions of normalizable string states in the background $\phi^{(0)}$ can  be obtained genus by genus through derivation with respect to the sources $J$ and by path integrating over the world sheet fields $X$ and the moduli space of Riemann surfaces. In principle, we can shift the background fields $\phi^{(0)}$ by (for instance) coherent states of the normalizable modes, and obtain a different background which is part of the same quantum theory \cite{Sen:1993kb}. The non-normalizable part of the fields of the theory however are given once and for all, and they specify the quantum theory of gravity at hand. The traditional definition of  perturbative first quantized string theory makes sense when the background as well as the perturbative excitations are on-shell.

We will study the generating function $Z[\phi^{(0)},J]$ and its derivatives in string theory in a three-dimensional space-time with a negative cosmological constant. The most frequently studied background in this category is the three-dimensional anti-de Sitter Poincar\'e patch with metric:
\begin{equation}
ds^2 = l^2 (\frac{dr^2}{r^2} + r^2 dx d \bar{x})  \, .
\label{AdS3Metric}
\end{equation}
The background is supported by a NS-NS three-form field strength $H_{(3)}$  which is proportional to the volume form on the $AdS_3$ space-time. The dilaton $\Phi$ is constant. 
The boundary metric  in the Poincar\'e patch  (\ref{AdS3Metric}) is flat. 
One of our  goals is to  understand  the bulk holographic counterpart to the distinction between a physical boundary theory with a standard energy-momentum tensor, and a topologically twisted boundary theory with an energy-momentum tensor which is trivial in cohomology. To make the distinction between energy-momentum tensors,  we need to study their coupling to the boundary metric. We therefore analyze  string theory on a broader set of backgrounds than the  $AdS_3$
space-time (\ref{AdS3Metric}) with flat boundary. 

In field theory, we  compute the energy-momentum tensor as the derivative of the action with respect to the metric. In our first quantized string theoretic approach, a good stand-in for the space-time effective action is the generator of correlation functions $Z$. Indeed, it has been argued  that the string effective action is closely related to the generating function of correlation functions (see e.g.  \cite{Tseytlin:1988rr} for a review). Thus, the boundary energy-momentum tensor will measure the change of the generating function $Z$ under perturbations of the boundary metric. 

Therefore, we are motivated to study string theory in a bulk space-time metric $G_{\mu \nu}$ which depends on a boundary space-time metric $g^{(0)}_{ij}$. To leading order in a radial expansion, we study strings propagating in the background space-time:
\begin{equation}
ds^2 = l^2 \left( \frac{dr^2}{r^2} + r^2 g_{ij}^{(0)}dx^i d x^j + O(r^0) dx^i d x^j \right)  \, .
\label{AdS3WithGenericBoundary}
\end{equation}
The boundary metric $g^{(0)}_{ij}$ is part of the non-normalizable boundary conditions in the bulk. It is important to note that its introduction will also lead to subleading terms in the metric since the total background must  remain on-shell.
Once we introduce the background boundary metric $g^{(0)}_{ij}$, the generating function $Z[g^{(0)}_{ij},J]$ becomes a functional of the metric. We can differentiate the generating function $Z$ with respect to the boundary metric to obtain a world sheet vertex operator that codes the space-time boundary energy-momentum tensor. This space-time energy-momentum tensor operator was determined in \cite{Giveon:1998ns,Kutasov:1999xu} for a flat boundary metric through slightly different means. A derivation that roughly matches the scheme outlined above can be found in \cite{Troost:2010zz}. We will generalize the derivation to the case of a generic conformally flat boundary.

Moreover, we wish to show that for $AdS_3$ backgrounds dual to $N=2$ superconformal field theories on the boundary, there is a definition of bulk backgrounds depending on boundary metrics $g^{(0)}_{ij}$ that leads to a topological boundary space-time energy-momentum tensor. Indeed, the boundary $N=2$ superconformal field theory can be twisted to a topological conformal field theory \cite{Eguchi:1990vz,Witten:1988xj}.
The  energy-momentum tensor of the untwisted boundary $N=2$ superconformal field theory  we denote by ${\cal T}$, its left-moving $U(1)_R$ current  by ${\cal J}^R$  and the left-moving $N=2$ supercurrents by ${\cal G}^{\pm}$.
The topological conformal field theory is defined by restricting observables to a zero mode supercharge ${{\cal G}}_0^+$ cohomology, and by observing that the twisted energy-momentum tensor
\begin{equation}
{\cal T}^{\text{top}}_{{x} {x}} = {\cal T}_{{x} {x}} + \frac{1}{2} \nabla_{{x}} {{\cal J}}^R_{{x}}
\end{equation}
is ${\cal G}_0^+$ exact as a consequence of the boundary $N=2$ superconformal algebra \cite{Eguchi:1990vz,Witten:1988xj}.
 Thus, part of our task is to show that there is a string theory generating function of correlation functions whose derivative with respect to the boundary metric gives rise to the  topological energy-momentum tensor ${\cal T}^{\text{top}}$. This is what we wish to demonstrate in this paper for backgrounds with a conformally flat boundary.

\section{The Three-Dimensional Stringy Geometries}
\label{Solutions}
In this section we dress standard three-dimensional solutions of Einstein gravity with a negative cosmological constant and arbitrary boundary metric \cite{Banados:1998gg} with a NS-NS three-form flux $H_{(3)}$ and a constant dilaton $\Phi$ such that they become solutions of the space-time equations of motion of string theory. We then include an extra circle direction that will model the  $U(1)_R$ symmetry of our boundary conformal field theory with extended supersymmetry. Next, we  twist the solution  to include a boundary metric dependent fibering of the circle over the anti-de Sitter space-time.
\subsection{The Stringy Standard Background}
Three-dimensional gravity with a negative cosmological constant permits the following  solution in Fefferman-Graham gauge \cite{Banados:1998gg}
\begin{eqnarray}
ds^2 &=& l^2 \left( \frac{dr^2}{r^2} + (r^2 g^{(0)}_{ij} + 
g^{(2)}_{ij}+ r^{-2} g^{(4)}_{ij}) dx^i dx^j \right) \, , \label{BanadosSolution}
\end{eqnarray}
where $g^{(0)}_{ij}(x^k)$ is a freely chosen boundary metric and the subleading terms in the metric are largely determined by the Einstein equations: 
\begin{eqnarray}
g^{(2)}_{ij} &=& -\frac{1}{2} R^{(0)} g^{(0)}_{ij} - \frac{4 G_N}{l}  \, T_{ij} \label{g2}
 \\
g^{(4)}_{ij} &=& \frac{1}{4} g^{(2)}_{ik} g^{(0)kl} g^{(2)}_{lj} \, . \label{g4}
\end{eqnarray}
We introduced the three-dimensional Newton constant $G_N$ and a conserved boundary energy-momentum tensor $T_{ij}$ which satisfies the trace anomaly $T^i_i=-c \, R^{(0)}/12$ with Brown-Henneaux central charge $c=3l/(2 G_N)$ \cite{Brown:1986nw}.
In the following we will consider a solution to string theory based on this general relativity background. We consider a string theory in its NS-NS sector, with a background metric $G_{\mu \nu}$, a NS-NS  three-form flux $H_{(3)}$ as well as a dilaton $\Phi$. For now, we will ignore the rest of the internal manifold of the string theory (which we do assume to be compact throughout).
To first order in the inverse string tension $\alpha'$ the equations of motion for these fields in  string theory read \cite{Polchinski:1998rq}:
\begin{eqnarray}
\alpha' R_{\mu \nu} + 2 \alpha' \nabla_\mu \nabla_\nu \Phi - \frac{\alpha'}{4}
H_{\mu \lambda \omega} {H_\nu}^{\lambda \omega} &=& 0
\nonumber \\
- \frac{\alpha'}{2} \nabla^\omega H_{\omega \mu \nu} + \alpha' \nabla^\omega \Phi H_{\omega \mu \nu}  &=& 0
\nonumber \\
\frac{c_{\text{matter}}-c_{\text{critical}}}{6} 
- \frac{\alpha'}{2} \nabla^2 \Phi + \alpha' \nabla_\omega \Phi \nabla^\omega \Phi - \frac{\alpha'}{24} H_{\mu \nu \lambda} H^{\mu \nu \lambda} &=& 0 
\, . \label{StringEquationsOfMotion}
\end{eqnarray}
We  take the dilaton $\Phi$ to be constant such that we have a constant string coupling. We moreover consider the metric solution (\ref{BanadosSolution}) which satisfies
\begin{equation}
R_{\mu \nu} = - \frac{2}{l^2} G_{\mu \nu} \, .
\end{equation}
A NS-NS three-form flux proportional to the volume form  saturates the first two equalities in the space-time equations of motion (\ref{StringEquationsOfMotion}):
\begin{equation}
H_{(3)} = \frac{2}{l} \sqrt{|G|} \, dx^\mu \wedge dx^\nu \wedge dx^\rho \, .
\end{equation}
Indeed, we then have that $H_{\mu \nu \rho} = 2/l \sqrt{|G|} \epsilon_{\mu \nu \rho}$ and since the space-time metric $G_{\mu \nu}$ is covariantly constant, we  satisfy the equations of motion. The last equation of motion in (\ref{StringEquationsOfMotion}) is also satisfied if we take into account the contribution of the flux to the total world sheet central charge, as is standard in $AdS_3$ string theory. We summarize the stringy standard background solution:
\begin{eqnarray}
ds^2 &=&  l^2 \left( \frac{dr^2}{r^2} + (r^2 g^{(0)}_{ij} + 
g^{(2)}_{ij}+ r^{-2} g^{(4)}_{ij}) dx^i dx^j \right)
\nonumber \\
H_{(3)} &=& \frac{2}{l} \sqrt{|G|}  \,dx^\mu \wedge dx^\nu \wedge dx^\rho 
\nonumber \\
\Phi &=& \text{constant} \, . \label{StringyStandardSolution}
\end{eqnarray}
This solution is standard in the sense that it is unique under the conditions that we keep the dilaton constant, we fix the metric to be of the form (\ref{BanadosSolution}) and we allow for NS-NS flux only, in the three directions of space-time at hand.\footnote{The assumption that the NS-NS-flux moves in lockstep with the metric propagates throughout the paper. A microscopic construction of the string theory background in terms of near horizon NS5-branes and fundamental strings is bound to obey it. }
We worked at the level of the gravitational approximation to string theory (as we will do in the rest of the paper), but it is important to realize that many considerations can be made exact in $\alpha'$. See e.g. \cite{Giveon:1998ns,Kutasov:1999xu,Maldacena:2001km,Troost:2010zz,Troost:2011ud}. While one is not able to solve the world sheet conformal field theory in all these backgrounds, performing  conformal perturbation theory in  generic deformations around Poincar\'e $AdS_3$ is within reach.
Finally, the variation of the generating function of correlation functions  with respect to the boundary metric component $\delta g^{(0)}_{xx}=h_{xx}$ near the  $AdS_3$ solution gives rise to the space-time energy-momentum component ${\cal T}^{xx}$ as described  in \cite{Giveon:1998ns,Kutasov:1999xu,Troost:2010zz}.

\subsection{The Twisted Background}
To obtain a topological string theory (in the sense that it will be independent of the boundary values of the metric), we  allow the boundary metric to couple to other degrees of freedom in the  theory \cite{Li:2019qzx}. Drawing inspiration from the topological twist of ordinary quantum field theories \cite{Witten:1988ze}, we expect that the boundary metric will also couple to the derivative of the R-symmetry current.

Indeed, to topologically twist we need a boundary conformal field theory with at least $N=2$ supersymmetry and therefore a $U(1)_R$ symmetry. The conditions on the bulk string theory in order to reach such boundary theories were carefully analyzed in \cite{Giveon:2003ku}. If we assume that the boundary conformal field theory has both a left and a right $N=2$ superconformal algebra, then a reasonably generic representation of backgrounds that allow for this boundary conformal symmetry are backgrounds with an extra circle direction. Thus, we  generalize the standard string background (\ref{StringyStandardSolution}) to include an extra circle.
When we have an extra direction in space-time, more general solutions to the equations of motion of string theory exist, including solutions that fibre the circle over the standard stringy solution constructed above.
To discuss these solutions, we first recall  the  dimensional reduction of a four-dimensional string effective action.

\subsubsection{Dimensional Reduction}
\label{DimensionalReduction}
We consider a four-dimensional space-time that  is the product of a three-dimensional space-time with negative cosmological constant and  a circle fibered over the locally $AdS_3$ factor. Again, we concentrate on the NS-NS sector. We pick a  dilaton $\Phi$ which is constant   and consider a four-dimensional metric $G^4_{MN}$ as well as a NS-NS two-form $B_{(2)}^4$ of the type:
\begin{eqnarray}
G_{MN}^4 dx^M  dx^N &=& G_{\mu \nu} dx^\mu dx^\nu + G_{44} (d \theta + A_\mu dx^\mu)^2
\nonumber \\
B_{(2)}^4 = \frac{1}{2} B_{MN} dx^M \wedge dx^N &=& \frac{1}{2} B_{\mu \nu} dx^\mu \wedge dx^\nu + 
 B_{\mu 4} dx^\mu \wedge d \theta \, ,
\end{eqnarray}
where the index $M$ can take the three-dimensional values $\mu$ as well as $4$. The fourth direction with coordinate $x^4=\theta$ is a compact direction and the angle $\theta$ is  identified modulo $2 \pi R$. When we dimensionally reduce the four-dimensional string effective action
\begin{eqnarray}
S_{\text{space-time}} &=& \frac{1}{2 \kappa_0^2} \int d^4 x (-G^4)^{\frac{1}{2}}
e^{-2 \Phi} (R+ 4 \nabla_\mu \Phi \nabla^\mu \Phi
- \frac{1}{12} H_{MNL} H^{MNL}) \, 
\end{eqnarray}
 on the circle,
we obtain the three-dimensional action  \cite{Polchinski:1998rq}:
\begin{eqnarray}
S_{\text{space-time}} &=& \frac{ 2 \pi R}{2 \kappa_0^2} \int d^3 x (-G)^{\frac{1}{2}} e^{- 2 \Phi_3}
\Big( R_3 - \partial_\mu \sigma \partial^\mu \sigma + 4 \partial_\mu \Phi_3 \partial^\mu \Phi_3 - \frac{1}{4} e^{2 \sigma} F_{\mu \nu} F^{\mu \nu} \nonumber \\
& & \qquad
- \frac{1}{12} ( \tilde{H}_{\mu \nu \lambda} \tilde{H}^{\mu \nu \lambda} +
3 e^{-2 \sigma} H_{4 \mu \nu} {H_4}^{\mu \nu}) \Big)
\end{eqnarray}
where we introduced the notations
\begin{eqnarray}
G_{44} &=& e^{2 \sigma} \, , \qquad
\Phi_3 = \Phi - \sigma/2 \, , 
\nonumber \\
\tilde{H}_{\mu \nu \lambda} &=& (\partial_\mu B_{\nu \lambda}-A_\mu H_{4 \nu \lambda})+\text{cyclic} \, .
\end{eqnarray}
If we pick both the dilaton $\Phi$ and the metric component $G_{44}=e^{2 \sigma}$ to be constant, then the action reduces to a three-dimensional gravity coupled to two Maxwell fields as well as an anti-symmetric two-form tensor:
\begin{eqnarray}
S_{\text{space-time}} &=& \frac{ 2 \pi R}{2 \kappa_0^2} \int d^3 x (-G)^{\frac{1}{2}} e^{- 2 \Phi_3}
(R_3  
- \frac{1}{12}  \tilde{H}_{\mu \nu \lambda} \tilde{H}^{\mu \nu \lambda} 
- \frac{1}{4} e^{2 \sigma} F_{\mu \nu} F^{\mu \nu} - \frac{1}{4} e^{-2 \sigma} H_{4 \mu \nu} {H_4}^{\mu \nu})  \, . \nonumber \\
& & \label{DimensionallyReducedAction} 
\end{eqnarray}
Thus, we have an almost standard three-dimensional Neveu-Schwarz-Neveu-Schwarz sector, as well as two Maxwell fields $A_\mu$ and $B_{4 \mu}$. 
In the following, we  set the metric component $G_{44}=1$
by rescaling the coordinate $\theta$. We furthermore  redefine the Maxwell fields into a new pair of $U(1)$ gauge fields:
\begin{eqnarray}
A^R &=& \frac{1}{2l} (A_\mu + B_{\mu4}   ) 
\nonumber \\
\bar{A}^R &=& \frac{1}{2l} (A_\mu - B_{\mu4}   ) \, .
\end{eqnarray}

\subsubsection{Degrees of Freedom}
An understanding of the dynamics of the gauge field fluctuations around the solution (\ref{StringyStandardSolution}) will be useful. To that end, we plug in the background NS-NS $H_{(3)}$ flux proportional to the volume form in
(\ref{DimensionallyReducedAction}). We concentrate on the fluctuations of the gauge fields $A^R$ and $\bar{A}^R$ only, and find the quadratic part of the effective action
\begin{equation}
S_{\text{gauge}}^{\text{quad}}
=  
 \frac{ 2 \pi R e^{-2 \Phi_3}}{2 \kappa_0^2} \int d^3 x  \left(
2l (A^R \wedge dA^R-\bar{A}^R \wedge d \bar{A}^R) -l^2 (F^R \ast F^R + \bar{F}^R \ast \bar{F}^R ) \right) \,  . \, \label{QuadraticEffectiveAction}
\end{equation}
The free part of the equation of motion of the gauge field $A^R$ reads:
\begin{equation}
d \ast F^R = -\frac{2}{l} dA^R \, . \label{FEOM}
\end{equation}
As a consequence, the gauge field $A^R$ can be split into two terms \cite{Andrade:2011sx}:
\begin{equation}
A^R = A^R_{\, \text{flat}} - \frac{l}{2} \ast F^R \, .
\label{GaugeFieldDecomposition}
\end{equation}
The first term consists of a flat connection $A^R_{\, \text{flat}} $. The second term  is proportional to the Hodge dual of the gauge invariant field strength and satisfies the equation of motion
\begin{equation}
l \ast d (A^R-A^R_{\, \text{flat}}) = - 2 (A^R-A^R_{\, \text{flat}})
\end{equation}
of a massive field dual to a $(2,1)$ primary operator \cite{Gukov:2004ym}. It is important to us that the flat part of the gauge field is dual to the current while its curvature is dual to a higher dimensional operator \cite{Gukov:2004ym}. Moreover, note that the flat gauge field $A^R_{\, \text{flat}} $ satisfies a first order differential equation and that we therefore expect to fix a single component of the flat gauge field  at infinity when solving the equations of motion.

\subsubsection{Flat Generalizations}
\label{FlatGeneralization}
The dimensionally reduced action (\ref{DimensionallyReducedAction}) makes it manifest that we can generalize the stringy standard solution (\ref{StringyStandardSolution}). Indeed, 
when the two Maxwell fields are flat, the NS-NS  equations of motion reduce to the three-dimensional equations of motion (\ref{StringEquationsOfMotion}).
We conclude that the stringy standard background (\ref{StringyStandardSolution}) can be augmented to a four-dimensional solution:
\begin{eqnarray}
\Phi &:& \text{constant}
\nonumber \\
G_{\mu \nu} &:& \text{locally} \, \, AdS_3 \, \, \text{with a non-trivial boundary metric}
\nonumber \\
H &:& \text{proportional to the volume form}
\nonumber \\
G_{4 4} &:& \text{constant}
\nonumber \\
{A}^R \,\, \text{and} \, \, \bar{A}^R &:& \text{flat}
 \, .
\end{eqnarray}

\subsubsection{The Explicit Boundary Gauge Field Dependence}
\label{ExplicitGaugeField}
Firstly, we consider a class of solutions that allows us to compute the boundary $U(1)_R$ current ${{\cal J}}^R$. The current couples to flat boundary gauge field fluctuations.
Thus, we add a flat boundary fluctuation of the gauge field to the background solution (\ref{StringyStandardSolution}). We parameterize the solution in terms of the fluctuation of the gauge field component $\delta A^R_{\bar{x}}$ and find from the flatness equation that the other component is given by: 
\begin{equation}
\delta A_x^R = \int^{\bar{x}} \partial_x \delta A_{\bar{x}}^R \, .   \label{GaugeFieldFluctuationSolution}
\end{equation}
 This is a simple example of how the solution takes on a non-local character when we parameterize it in terms of boundary fluctuations that couple directly to conserved currents. It is important to note that parameterising the theory in terms of a given boundary component is tantamount to adding a  particular boundary term to the action \cite{Kraus:2006wn}. The resulting action should have an energy bounded from below. This requirement fixes the boundary component to be chosen in terms of the sign of the Chern-Simons level of the quadratic effective action (\ref{QuadraticEffectiveAction}) for the gauge fields \cite{Kraus:2006wn}. We have chosen our boundary component accordingly.

\subsubsection{The Explicit Boundary Metric Dependence}
\label{Non-LocalMetric}

In the following, we choose to perturb the boundary metric around a conformally flat  metric. Conformally flat background metrics have the advantage of allowing for non-trivial boundary curvature $R^{(0)}$, while still retaining some of the simplicity of the background with a flat boundary metric.\footnote{The curvature is crucial, to give but one example, to understand  the boundary theory on a two-sphere. Indeed, while the boundary theory we are aiming for may be locally independent of the metric, it still  depends on topological curvature invariants like the Euler number of the boundary Riemann surface.}
Thus, we firstly note that the equation (\ref{g2}) for the subleading metric perturbation combined with the trace condition on the energy-momentum tensor permits a closed form local solution for the conformally flat boundary metric:
\begin{equation}
g^{(0) \, \text{conf. flat}}_{ij} dx^i dx^j = e^{2 \omega} dx d \bar{x} \, .
\label{ConformallyFlatMetric}
\end{equation}
In the following we  work with a background boundary metric which is a small perturbation  of the conformally flat metric, and we shall solve for the bulk metric explicitly.  In this generalization, non-local terms  appear.

Thus, we wish to compute the explicit subleading metric dependence of the standard solution (\ref{StringyStandardSolution}) for a boundary metric of the form:\footnote{We often set the cosmological constant length scale $l=1$ from now on.}
\begin{eqnarray}
ds^2 &=& \frac{dr^2}{r^2}+r^2 e^{2\omega} dx d\bar{x}+r^2 h_{xx}dxdx
\nonumber\\
& &  +(\partial_x^2\omega-(\partial_x\omega)^2)dxdx 
 +2\partial_x\partial_{\bar{x}}\omega dxd\bar{x}+(\partial_{\bar{x}}^2\omega-(\partial_{\bar{x}}\omega)^2)d\bar{x}d\bar{x}
\nonumber\\
& & +\delta g^{(2)}_{xx} dxdx 
 +2\delta g^{(2)}_{x\bar{x}}dxd\bar{x}+\delta g^{(2)}_{\bar{x}\bar{x}} d\bar{x}d\bar{x}
\nonumber\\
&& + O( r^{-2} ) \, .
\label{ConformalCase}
\end{eqnarray}
As advertised, we have added a conformal factor $e^{2 \omega}$ and have explicitly solved for the $O(r^0)$ subleading term as a  function of the conformal factor $\omega$. On top of this conformally flat boundary metric, we have added a leading $O(r^2)$ perturbation $h_{xx}$ of the boundary metric component $g^{(0)}_{xx}$. Our task is to explicitly solve for the dependence of the subleading terms $\delta g^{(2)}_{ij}$ as a functional of the leading perturbation $h_{xx}$. 
Solving Einstein's equations to leading order in the metric perturbation $h_{xx}$ leads to the equalities:
\begin{eqnarray}
\delta g^{(2)}_{xx} &=& -\frac{1}{2}e^{-2\omega}
(\partial_x\partial_{\bar{x}}h_{xx}
-2\partial_x\partial_{\bar{x}}\omega h_{xx}
-4\partial_x\omega\partial_{\bar{x}}h_{xx})\, ,\\
\delta g^{(2)}_{x\bar{x}} &=& -\frac{1}{2}e^{-2\omega}
[\partial_{\bar{x}}^2 h_{xx}
-2\partial_{\bar{x}}\omega\partial_{\bar{x}}h_{xx}
-2(\partial_{\bar{x}}^2\omega-(\partial_{\bar{x}}\omega)^2)h_{xx}]\, ,\\
\delta g^{(2)}_{\bar{x}\bar{x}} &=& -\frac{1}{2}\int^x \partial_{\bar{x}}^3(e^{-2\omega}h_{xx}) \, .
\end{eqnarray}
The exact subleading $O(r^{-2})$ metric dependence on the perturbation is then easily obtained by plugging this perturbation for $g^{(2)}_{ij}$ into the consequence (\ref{g4}) of Einstein's equations.

To understand the string theory background to this order, it is  useful to be more explicit about the NS-NS two-form potential $B_{(2)}$.
We note that the square root of the determinant of the three-dimensional metric $G_{\mu \nu}$ (\ref{BanadosSolution}) equals:
\begin{equation}
\sqrt{G} =  r \sqrt{g^{(0)}} (1+ \frac{g^{(0)ij} g_{ji}^{(2)}}{2 r^2} + \frac{1}{4 r^4} \frac{g^{(2)}}{g^{(0)} }) \, .
\end{equation}
We can thus radially integate the NS-NS three-form to obtain the two-form gauge potential:
\begin{eqnarray}
B_{(2)} &=& \sqrt{ g^{(0)}} (r^2  + \log r   \, g^{(0)ij} g_{ji}^{(2)}-\frac{1}{4 r^2} \frac{g^{(2)}}{{ g^{(0)}} } )d x^i \wedge dx^j \, .  \label{IntegratedBField}
\end{eqnarray}
We keep our string background on the mass shell, and compute the perturbation of the NS-NS two-form potential as a consequence of the metric perturbation $h_{xx}$ to zeroth (or logarithmic) order in the radial coordinate $r$.
We note that the metric perturbation $h_{xx}$ does not change (the square root of the absolute value of) the determinant
of the boundary metric $g_{ij}^{(0)}$ -- it remains $\sqrt{ g^{(0)}}=\frac{1}{2}e^{2 \omega}$.
Furthermore, 
we recall that the trace of the second order of the boundary metric equals half the boundary Ricci scalar, $g^{(0)ij} g_{ji}^{(2)}=- R^{(0)}/2$, 
such that we need the dependence of the  Ricci-scalar on the perturbation $h_{xx}$ to first order:
\begin{eqnarray}
R^{(0)} &=& 4 e^{-4 \omega} (-2 \partial_{\bar{x}} h_{xx}
\partial_{\bar{x}} \omega+ \partial_{\bar{x}}^2 h_{xx}
-2 e^{2 \omega} \partial_{{x}} \partial_{\bar{x}} \omega) \, .  \label{RicciScalarPerturbation}
\end{eqnarray}
Combining equations (\ref{IntegratedBField}) and (\ref{RicciScalarPerturbation}), we  know explicitly the NS-NS two-form potential $B_{(2)}$ to first order in the metric perturbation $h_{xx}$.

\subsubsection{The Asymptotic Twisted Generalization}
\label{TwistedGeneralization}
Finally, we generalize the solution (\ref{StringyStandardSolution}) to the background central to our intent. We introduce a dependence of the  $U(1)_R$ gauge field on the asymptotic boundary metric $g^{(0)}_{ij}$ in order to couple the boundary metric non-trivially to the R-current. We draw inspiration from the analogous exercise  performed in supergroup Chern-Simons theory  \cite{Li:2019qzx} as well as from the literature on topological quantum field theories  \cite{Witten:1988ze}.
We wish to introduce a coupling $({\cal J}^R \pm {\bar {\cal J}}^R)_\mu
\omega^\mu$ between the R-currents and the spin connection one-form $\omega=\omega_\mu dx^\mu$ on the boundary, where the relative sign depends on the twist we perform.

In the following, we work near conformally flat boundaries. In conformally flat backgrounds, the coordinates $x$ and $\bar{x}$ parameterize light-cone directions in Lorentzian signature. The boundary R-currents remain chiral, and each current only couples to a single component of the spin connection one-form. Thus, the gauge fields  $A^{R}$ and $\bar{A}^{R}$ near conformally flat backgrounds are expected to have the boundary profile
\begin{equation}
A^R_{\bar x} =  - \frac{i}{4} \omega^{+-}_{\bar x}
\, , \qquad \quad 
\bar{A}^R_x =  \mp  \frac{i}{4} \omega^{+-}_x
\label{TwistedBoundaryAsymptotics}
\end{equation}
where $\omega^{+-}$ is the  spin connection one-form associated to the boundary metric $g^{(0)}_{ij}$.\footnote{We explicitly indicate the upper indices on the single component of the one-form spin connection in order to avoid a clash of notation with the conformal factor $\omega$.} We choose the perturbed zweibeins $e^{\pm}$:
\begin{equation}
e^+ = e^{\omega} dx \, , \quad  \qquad  e^- = e^{\omega} d \bar{x} + e^{-\omega} h_{xx} dx \, .
\end{equation} 
When we perform a metric perturbation $h_{xx}$ on top of the conformal background, the response of the spin connection  is chiral and equal to
\begin{equation}
\delta \omega_x^{+-} = -2 e^{-2 \omega} \partial_{\bar x} h_{xx} 
= -2 \nabla_{\bar x} (g^{\bar{x} x} h_{xx}) \, ,
\qquad \delta \omega_{\bar x}^{+-} = 0 \, .
\end{equation} 
Thus, the response of the gauge field component $\bar{A}^R_{\bar x}$ to the perturbation equals
\begin{eqnarray}
\delta \bar{A}^R_x &=& \pm \frac{i}{2} \nabla_{\bar x} (g^{\bar{x} x} h_{xx}) \label{TwistedSolution} \, ,
\end{eqnarray}
while the gauge field $A^R$ does not vary.
We will think of the perturbation as pertaining to the flat gauge field $\bar{A}^R_{\text{flat}}$ introduced in equation (\ref{GaugeFieldDecomposition}) which couples to the boundary R-current. The other component of the  gauge field guarantees that it indeed remains flat, as in equation (\ref{GaugeFieldFluctuationSolution}).

\section{The Space-Time Energy-Momentum  Operator}
\label{Tensors}\label{Operators}\label{Energy-Momentum}
 In this section we use the perturbed background metrics and gauge potentials of section \ref{Solutions} and plug them into the generating function  reviewed in section \ref{GeneratingFunction}. We then differentiate with respect to the metric and gauge field perturbations to obtain the world sheet vertex operators that correspond to the boundary space-time energy-momentum tensor and R-current in backgrounds with conformally flat boundary. We prove that the twisted string theory background leads to a topological  energy-momentum tensor. 
 \subsection{The Energy-Momentum}
 The world sheet vertex operator that represents the physical energy-momentum tensor of the boundary conformal field theory in $AdS_3$ string theory was derived in
\cite{Giveon:1998ns,Kutasov:1999xu} for a flat boundary. In the following, we compute the energy-momentum tensor  from the perspective of the generating function $Z[g^{(0)}_{ij}]$ of correlation functions, for all conformally flat backgrounds. 
Firstly, we define a world sheet vertex operator ${\cal T}_{\bar{x} \bar{x}}$ as the derivative of the generating function of correlation functions $Z[g^{(0)}_{ij}]$ with respect to the boundary metric perturbation $\delta g^{(0)}_{xx}=h_{xx}$ relative to the conformally flat background metric (\ref{ConformallyFlatMetric}). We use the  world sheet action $S$,
\begin{equation}
S = \frac{1}{2 \pi \alpha'} \int d^2 z (G_{MN}+B_{MN}) \partial X^N \bar{\partial} X^M \, ,
\end{equation}
the metric $G_{MN}$ and the NS-NS two-form $B_{(2)}$ that we obtained in section \ref{Solutions}, as well as the definition (\ref{DefinitionGeneratingFunction}) of the generating function $Z$. These formulas enable us to functionally differentiate the  world sheet vertex operator that represents the physical boundary energy-momentum tensor component ${\cal T}_{\bar{x}\bar{x}}$, to subleading order in the $r^{-2}$ expansion:
\begin{eqnarray}
{\cal T}_{\bar{x}\bar{x}}(x',\bar{x}') &=&  \frac{l^2 e^{2 \omega(x')}}{ \alpha'} \int d^2 z \Big[ r^2 \delta^{(2)}(x-x') \partial x \bar{\partial} x 
\nonumber \\
& &  -\frac{1}{2}
[(e^{-2\omega}\partial_x {\partial}_{\bar{x}}  \delta^{(2)}(x-x'))
-2\partial_x {\partial}_{\bar{x}}\omega \delta^{(2)}(x-x')
-4\partial_x \omega {\partial}_{\bar{x}} ( \delta^{(2)}(x-x'))] \partial x \bar{\partial} x \,   \nonumber \\
& &  -\frac{1}{2}
[\partial_{\bar{x}}(e^{-2\omega(x)}\partial_{\bar{x}}\delta^{(2)}(x'-x)) -2 e^{-2\omega}({\partial}_{\bar x}^2\omega-({\partial}_{\bar x}\omega)^2)\delta^{(2)}(x-x')] (\partial x \bar{\partial} \bar{x} +
\partial \bar{x} \bar{\partial} {x}) \, \nonumber \\
&&  -\frac{1}{2} \int^x  {\partial}^3_{\bar{x}} (e^{-2\omega}\delta^{(2)}(x-x'))
{\partial} \bar{x} \bar{\partial} \bar{x} 
\, \label{Energy-MomentumTensor} \\
&& - \log r   \partial_{\bar{x}} ( e^{-2 \omega}  \partial_{\bar{x}} ( \delta^{(2)}(x-x')))
(\partial \bar{x} \bar{\partial}x - \partial x \bar{\partial} \bar{x})  \Big] \, . \nonumber 
\end{eqnarray}
This is the space-time boundary energy-momentum tensor component ${\cal T}_{\bar{x}\bar{x}}$ for locally $AdS_3$ backgrounds with a conformally flat boundary. There is a leading term of order $r^2$, a subleading symmetric term independent of the radius, and a logarithmic term in the radius that originates from the anti-symmetric tensor.\footnote{The logarithmic behaviour is familiar from the linear dilaton term in the action of a long string \cite{Seiberg:1999xz}.} When we restrict our general expression to the case of a flat boundary metric (namely we put $\omega=\text{constant}$) then the space-time energy-momentum tensor agrees with the expression for the tensor found in \cite{Giveon:1998ns,Kutasov:1999xu} up to an anti-symmetric tensor gauge transformation.\footnote{
This is demonstrated in detail in Appendix \ref{Reduction}.} 
The equations of motion can be used to simplify the expression (\ref{Energy-MomentumTensor}). One can show that the  term of order $O(r^2)$ in the energy-momentum tensor collapses to a term of order $O(r^0)$ on shell and that  the non-local term vanishes to  order $O(r^0)$.\footnote{See Appendix \ref{Conservation}.}

Secondly, in a conformally flat background the boundary can be curved and  the ${\cal T}_{x \bar{x}}$ component of the energy-momentum tensor can also be non-zero. Because the metric variation $\delta h_{x \bar{x}}$ amounts to varying the conformal factor $\omega$ in the metric, the trace of the energy-momentum tensor is easier to compute. We find that the component ${\cal T}_{x\bar{x}}$ is given by
\begin{equation}
{\cal T}_{x\bar{x}} = \frac{e^{4\omega}}{4} {\cal T}^{x\bar{x}} 
= \frac{e^{4\omega}}{4}\frac{1}{2}\frac{4\pi\delta S}{\sqrt{g^{(0)}}\delta h_{x\bar{x}}}
= \frac{\pi\delta S}{\delta\omega} \, .
\end{equation}
The leading order in the vertex operator ${\cal T}_{x\bar{x}}$ equals
\begin{eqnarray}
{\cal T}_{x\bar{x}} &=& 
\frac{l^2}{\alpha'}\int d^2z
\Big[\partial_x\partial_{\bar{x}}\delta^{(2)}(x'-x)\partial x\bar{\partial}\bar{x}- \log r \partial_x \partial_{\bar{x}} \delta^{(2)}(x'-x)\partial x\bar{\partial} \bar{x} \Big] \nonumber \\
&=&
\frac{l^2}{\alpha'}\int d^2z
\Big[\partial_x\partial_{\bar{x}}\delta^{(2)}(x'-x)\partial x\bar{\partial}\bar{x}+\partial_x\delta^{(2)}(x'-x)\partial x\bar{\partial} \log r\Big] \, .
\end{eqnarray}
Upon partial integration, the first term is subleading, and the second term can be molded into a more recognisable form using the equations of motion:
\begin{equation}
{\cal T}_{x\bar{x}}  =  \frac{l^2}{\alpha'}\int d^2z
\partial \delta^{(2)}(x'-x) \bar{\partial}\log r 
 =  -
 \frac{l^2}{\alpha'}\int d^2z
\partial \delta^{(2)}(x'-x) \bar{\partial} \omega
 =  \frac{l^2}{\alpha'}\int d^2z
\delta^{(2)}(x'-x)\partial\bar{\partial}\omega \, .
\nonumber
\end{equation}
The trace of the energy momentum tensor is therefore equal to
\begin{eqnarray}
4e^{-2\omega(x')}{\cal T}_{x\bar{x}} &=& -\frac{l^2  R^{(0)}}{2 \alpha'}  \int d^2z \delta^{(2)}(x'-x)\partial x \bar{\partial}\bar{x}
\, .
\end{eqnarray}
We recall the space-time central charge operator $C^{\, \text{st}}$ from   \cite{Kutasov:1999xu}:
\begin{eqnarray}
C^{\, \text{st}} &=&  6 \frac{ l^2}{\alpha'} \int d^2z \delta^{(2)}(x'-x)\partial x \bar{\partial}\bar{x} \, ,
\end{eqnarray}
and conclude that the trace of the energy momentum tensor ${\cal T}$ satisfies the operator relation
\begin{equation}
{\cal T}_i^i = 
-\frac{C^{\, \text{st}}}{12}R^{(0)}
\, .
\end{equation}
Thus, we have obtained all the components of the energy-momentum tensor.\footnote{ The component ${\cal T}_{xx}$ follows from a parity transformation applied to ${\cal T}_{\bar{x} \bar{x}}$.} Finally, the proof that the energy-momentum tensor in the conformally flat background is conserved, namely that the operator equation $\nabla^\mu {\cal T}_{\mu \nu}=0$ is valid, is given in appendix \ref{Conservation}.

\subsection{The R-Symmetry Current}
The R-current ${\cal J}^R$ couples to the perturbation of the boundary gauge field. We have prepared the ground for deriving the space-time R-current in subsection \ref{ExplicitGaugeField}. 
Using the explicit expression (\ref{GaugeFieldFluctuationSolution}) for the gauge field fluctuation and plugging it into the string world sheet action, and functionally differentiating, we find the R-current component:
\begin{eqnarray}
{\cal J}_x^R (x^\prime,\bar{x}^\prime) 
&=&  \frac{2 i l}{\alpha'}
\int d^2 z [\delta^{(2)} (x-x^\prime) \bar{\partial} \bar{x}  \partial \theta + \frac{\delta A_x^R(x,\bar{x})}{\delta A^R_{\bar{x}}(x^\prime,\bar{x}^\prime)} \bar{\partial} x \partial \theta]
\nonumber \\
&=& \frac{2 i l}{\alpha'}
\int d^2 z [\delta^{(2)} (x-x^\prime) \bar{\partial} \bar{x}  \partial \theta + \int^{\bar{x}} \partial_x 
\delta^{(2)}(x-x')
\bar{\partial} x \partial \theta]
\, .
\end{eqnarray}
We find a single current component since the flat gauge field is parameterized in terms of a single component of the boundary gauge field.\footnote{To match onto the analysis of \cite{Giveon:2003ku} with a flat boundary but a more general world sheet theory, one identifies their world sheet current $J^0$ with our $J^0 = i \sqrt{2/\alpha'} \partial \theta$, ignoring fermions throughout.} 
We can show that the current component ${\cal J}_x^R $ is holomorphic in conformally flat backgrounds,
$\partial_{\bar x}
{\cal J}_x^R =0$, up to contact terms in the quantum theory. Indeed, since the
gauge field to which the current couples is flat, we have:
 \begin{equation}
\partial_{\bar{x}} \frac{\delta A^R_x(x,\bar{x})}{\delta A^R_{\bar{x}}(x^\prime,\bar{x}^\prime)} = \partial_x \delta^{(2)} (x^\prime-x)
 \, .
  \end{equation}
We can use this equation, the chain rule for functional derivation, as well as the chirality of the world sheet current $\partial \theta$,
to show that
 \begin{equation}
 \partial_{\bar{x}^\prime} {\cal J}_x^R (x^\prime,\bar{x}^\prime) = 
  -\frac{2il}{\alpha^\prime} \int d^2 z \bar{\partial } (\delta^{(2)} (x^\prime-x)) \partial \theta \, 
\end{equation}
 equals zero after partial integration.\footnote{Our reasoning generalizes   the derivation in \cite{Kutasov:1999xu} to conformally flat backgrounds.}${}^{,}$\footnote{We assume a vanishing gauge field background in the proof.}

\subsection{The Topological Energy-Momentum Tensor}

\label{Twisted Energy-Momemtum}

Recall that we work under the assumption that we have a string theory with space-time boundary $N=2$ superconformal symmetry. Topologically twisting that superconformal field theory gives rise to a topological energy-momentum tensor of the form \cite{Eguchi:1990vz,Witten:1988xj}
\begin{equation}
{\cal T}^{\text{top}}_{xx} = {\cal T}_{xx} + \frac{1}{2} \nabla_x {\cal J}^R_x \, . \label{Ttop}
\end{equation} 
The topological energy-momentum tensor is ${\cal G}_0^+$ exact because it follows from the boundary $N=2$ superconformal algebra that:
\begin{equation}
{\cal T}^{\text{top}}_{xx} = [ {\cal G}^+_0,  {\cal G}^- ] \, .
\label{Exact}
\end{equation}
The boundary twisted $N=2$ superconformal theory has correlators that do not depend on states that are exact in ${\cal G}^+_0$ cohomology, and therefore they do not depend on boundary metric perturbations -- they are topological. In this subsection, we finally wish to prove that the twisted boundary conditions that we imposed on our string theory when we have a non-trivial boundary metric, lead to a world sheet vertex operator expression for the space-time energy-momentum tensor ${\cal T}^{\text{top}}$ which verifies the relation (\ref{Ttop}) and therefore indeed gives rise to a bulk string theory that is independent of the boundary metric.

Again, we differentiate the generating function of correlation functions $Z[g^{(0)}_{ij}]$ with respect to the boundary metric component $h_{xx}$, but now with twisted boundary asymptotics (\ref{TwistedBoundaryAsymptotics}). We find on the one hand the terms that 
we identified as the physical boundary energy-momentum tensor ${\cal T}_{\bar{x} \bar{x}}$ (\ref{Energy-MomentumTensor}), and a few extra terms originating in the part of the action which depends on the gauge fields:
\begin{equation}
{\cal T}^{\text{top}}_{\bar{x} \bar{x}} = {\cal T}_{\bar{x} \bar{x}}+ 2\pi e^{2\omega}\int d^2x^\prime 
\frac{\delta S}{\delta\bar{A}_{\bar x}(x^\prime,\bar{x}^\prime)}
\frac{\delta\bar{A}_{\bar x}(x^\prime,\bar{x}^\prime)}{\delta h_{xx}} ={\cal T}_{\bar{x} \bar{x}} - ie^{2\omega}\int d^2x^\prime 
{\bar{{\cal J}}}^R_{\bar{x}}(x^\prime,\bar{x}^\prime)\frac{\delta\bar{A}_x(x^\prime,\bar{x}^\prime)}{\delta h_{xx}}\, . \label{ExtraTerms}
\end{equation}
Using the gauge field variation \eqref{TwistedSolution}, one finds the topological energy-momentum tensor
\begin{eqnarray}
{\cal T}^{\text{top}}_{\bar{x} \bar{x}} &=& {\cal T}_{\bar{x} \bar{x}} \mp \frac{1}{2}e^{2\omega}\partial_{\bar{x}}(e^{-2\omega}\bar{\cal J}^R_{\bar{x}})\nonumber\\
&=& {\cal T}_{\bar{x} \bar{x}}
\mp \frac{1}{2} \nabla_{\bar{x}} \bar{\cal J}^R_{\bar{x}}\, ,   \label{MainPoint}
\end{eqnarray}
in the conformally flat background (\ref{ConformallyFlatMetric}).
The extra terms  agree with half the covariant derivative of the boundary R-current, thus proving the desired formula (\ref{Ttop}) for our twisted string theory.\footnote{We have proven the anti-holomorphic counterpart of the formula. The holomorphic version is proven analogously.
}${}^{,}$\footnote{ In Appendix \ref{TwistedTrace} we briefly discuss how the trace of the energy-momentum tensor is modified in the twisted theory.} The final result is closely tied to our choice of twisted asymptotic boundary conditions (\ref{TwistedBoundaryAsymptotics}). Indeed, the final equation follows from the twisted boundary condition combined with the definition of the R-symmetry current. Our analysis has shown that we can embed the whole of this reasoning in on-shell string theory.

The expression (\ref{MainPoint}) for the boundary energy-momentum tensor, combined with the boundary $N=2$ superconformal algebra, guarantees an anomalous R-charge conservation rule for the correlators in the topologically twisted string theory. It will be interesting to analyze its consequences, for instance by computing twisted bulk string theory correlation functions from first principles.

\section{Conclusions}
\label{Conclusions}
We analyzed string theory  in locally three-dimensional anti-de Sitter space-time  with a general boundary metric. We computed the world sheet vertex operator that represents the space-time energy-momentum tensor for any background with a conformally flat boundary. Furthermore, in the presence of a boundary $N=2$ superconformal algebra, we demonstrated that there exists an on-shell twist of the boundary conditions on the bulk string theory such that the boundary energy-momentum tensor operator  becomes topological. Indeed, we know that the  space-time energy-momentum tensor is topological since on the one hand, the space-time $N=2$ superconformal algebra can be derived from the bulk $AdS_3$ string theory, and on the other hand, the symmetry algebra implies that the topological energy-momentum tensor is ${\cal G}_0^+$ exact.

More work needs to be done to understand the bulk topological string theory well. Firstly, one would like to compute the ${\cal G}_0^+$ cohomology directly from the bulk perspective (as was done in the Chern-Simons supergravity in \cite{Li:2019qzx}). One can be hopeful that this is possible to all orders in the string length over the curvature radius, in a generic conformal field theory describing a $N=2$ superconformal string background. The fact that world sheet and space-time chiral primaries are closely related should help in this enterprise. Secondly, 
one needs a  first principle approach to the topological string world sheet correlation functions in a fixed $AdS_3$ background that takes into account (for instance) the space-time anomalous R-charge conservation. For  this phenomenon, the coupling of the string excitations to the scalar curvature of the boundary metric is paramount, and that coupling plays a key role in the topological correlation functions already  for a boundary sphere.  The conformally flat background with non-zero conformal factor $\omega$ corresponding to a two-sphere is thus an ideal playground to start computing topological string theory correlation functions. A  goal is to compute these correlation functions to all orders in the genus expansion in order to match known boundary correlators at finite central charge. 

While the $N=2$ boundary superconformal algebra forms the backdrop to our reasonings, we have only considered the bosonic subsectors of our world sheet and space-time string theory in  explicit calculations. Clearly, it would be desirable to extend our calculations to both world sheet and space-time fermions. The formulation of our asymptotic boundary conditions is such that the extra terms that are generated in this process will also satisfy the twist captured by equations (\ref{ExtraTerms}) and (\ref{MainPoint}).

It will certainly be interesting in these projects to keep the comparison with the approaches of \cite{Sugawara:1999fq,Rastelli:2005ph,Eberhardt:2019ywk,Eberhardt:2020akk,Costello:2020jbh} in mind. For instance, it would be useful to establish whether the topologically twisted theory we approach through twisted boundary conditions is equivalent to the theory based on the generalized BCOV topological string theory \cite{Bershadsky:1993cx,Costello:2012cy,Costello:2020jbh}. Likewise, it may be instructive to twist the superstring theory at curvature radius $\sqrt{\alpha'}$ to compare it with (topological aspects of) the physical theory
\cite{Eberhardt:2019ywk}. 

In our string theory background, we chose to lock the three-form NS-NS flux $H_{(3)}$ to the metric $G_{\mu \nu}$ while keeping the dilaton constant. The constant dilaton is a feature of near horizon limits of NS5-branes de-singularised by smeared fundamental strings. Our choice is likely to agree with a desired microscopic string theory construction of the $AdS_3$ backgrounds with non-trivial boundary metric and boundary topology. 

We recall that in the effective Chern-Simons-Maxwell theory for the R-symmetry gauge fields, one obtains (canonical) massive vector fields that are the Hodge duals of the (left and right) R-symmetry field strengths.  It might be enlightening  to determine  a bulk solution including the massive component of the gauge field e.g. using Fefferman-Graham perturbation theory on the full equations of motion of the action (\ref{DimensionallyReducedAction}), and to turn on vacuum expectation values for the dual operators of dimensions $(2,1)$ and $(1,2)$.

We concentrated on $AdS_3$ gravity with standard boundary conditions, standard twist, and NS-NS flux.
It is also possible to  study more general boundary conditions for  $AdS_3$ gravity  such as those considered in \cite{Compere:2013bya,Troessaert:2013fma,Grumiller:2016pqb}, and their extended supersymmetric counterparts.\footnote{As a preliminary exercise, one may wish to determine which subset of boundary conditions are consistent with quantum gravity.}  Again, one can  perform the twist by picking the boundary conditions for the gauge fields such that the gauge field dependent part in the action takes the form of a coupling between the R-current and the boundary spin connection. Another variation consists in introducing the background gauge field in only one of the two equations in (\ref{TwistedBoundaryAsymptotics}). One  obtains the shift of the energy momentum tensor as in equation (\ref{MainPoint}) only for the holomorphic or the anti-holomorphic component. This corresponds to a half-twist \cite{Witten:1991zz} on the boundary. Finally, one can also apply our logic to the asymptotic symmetry algebra in the presence of Ramond-Ramond or mixed fluxes
\cite{Ashok:2009jw} and obtain a bulk topological theory in that manner.

While we concentrated on the AdS/CFT duality in three dimensions, the conceptual approach we laid out transposes in a straightforward manner to other dimensions. A universe of further developments is accessible in this manner.

Still more generally, we stress that the goal of proving topological subsectors of AdS/CFT correspondences is worth pursuing in the light of the central role of the correspondence in the broad debate on  properties of unitary quantum theories of gravity.

\section*{Acknowledgments}
It is a pleasure to thank our colleagues for creating a stimulating research environment.

\appendix
\section{A Reduction to the Literature}
\label{Reduction}
In this appendix, we demonstrate that our energy-momentum tensor (\ref{Energy-MomentumTensor}) reduces to the energy-momentum tensor of \cite{Giveon:1998ns,Kutasov:1999xu}
when restricted to a flat boundary with conformal factor $\omega=0$.
To compare to \cite{Giveon:1998ns,Kutasov:1999xu}, we change coordinates:
\begin{eqnarray}
r & \rightarrow & e^{\phi}
\nonumber \\
x & \rightarrow & \gamma
\nonumber \\
\bar{x} & \rightarrow & \bar{\gamma} \, . \label{NewCoordinates}
\end{eqnarray}
Furthermore, the exact conformal field theory approach of \cite{Giveon:1998ns,Kutasov:1999xu} matches naturally with a different gauge choice for the NS-NS two-form $B_{(2)}$. We have the NS-NS three-form flux $H_{(3)}$
\begin{eqnarray}
 H_{(3)}
&=& ( 2 e^{2 \phi} \sqrt{g^{(0)}} - \partial_{{\gamma}}^2 h_{\bar{\gamma} \bar{\gamma}}+\dots)
d \phi \wedge  d \gamma \wedge d \bar{\gamma} \, ,
\end{eqnarray}
and we can choose a gauge in which
\begin{eqnarray}
B_{\gamma\bar{\gamma}} &=& e^{2\phi}\sqrt{g^{(0)}}+O(e^{-2\phi}) \nonumber \\
B_{\bar{\gamma}\phi} &=& -\partial_{\gamma}h_{\bar{\gamma}\bar{\gamma}} +O(e^{-2\phi})\, . \label{CFTGauge}
\end{eqnarray}
Using the coordinates (\ref{NewCoordinates}), the expression for the energy-momentum tensor (\ref{Energy-MomentumTensor}) adapted to this gauge and to the case $\omega=0$, barred appropriately, reads:
\begin{eqnarray}
{\cal T}_{xx}(x,\bar{x}) &=&  \frac{l^2}{\alpha'} \int d^2 z \Big[ e^{2 \phi} \delta^{(2)}(\gamma-x) \partial \bar{\gamma} \bar{\partial}\bar{\gamma} 
  -\frac{1}{2}
\partial_x {\partial}_{\bar{x}} ( \delta^{(2)}(\gamma-x))\partial
\bar{\gamma} \bar{\partial} \bar{\gamma}
 \,   \nonumber \\
& &  -\frac{1}{2}
{\partial}_x^2 ( \delta^{(2)}(\gamma-x))
 (\partial \gamma \bar{\partial} \bar{\gamma} +
\partial \bar{\gamma} \bar{\partial} {\gamma})   -\frac{1}{2} \int^{\bar{x}} {\partial}^3_x \delta^{(2)}(\gamma-x)
{\partial} \gamma \bar{\partial} \gamma 
\, \ \\
&&  +\partial_x (\delta^{(2)}(\gamma-x))
\bar{\partial}\bar{\gamma}\partial\phi +O(e^{-2 \phi})
\Big] \, . \nonumber 
\end{eqnarray}
The equations of motion imply that we have the order estimates
\begin{equation}
\partial \bar{\gamma} = O(e^{-2 \phi})
\, , \qquad 
\bar{\partial} \gamma = O(e^{-2 \phi}) \, .
\end{equation}
We can use these estimates to neglect a number of terms in the energy-momentum tensor:
\begin{eqnarray}
{\cal T}_{xx}(x,\bar{x}) & \approx & \frac{l^2}{\alpha'}
\int d^2 z \Big[ e^{2 \phi} \delta^{(2)}(\gamma-x) \partial \bar{\gamma} \bar{\partial}\bar{\gamma} 
-\frac{1}{2}
{\partial}_x^2 ( \delta^{(2)}(\gamma-x))\partial \gamma \bar{\partial} \bar{\gamma}
+\partial_x (\delta^{(2)}(\gamma-x))
\bar{\partial}\bar{\gamma}\partial\phi
\Big] . \nonumber
\end{eqnarray}
Using a standard formula for the $\delta$-function, we find:
\begin{eqnarray}
{\cal T}_{xx}(x,\bar{x}) &=& \frac{l^2}{\alpha'}\int d^2z \Big[ e^{2 \phi}  \partial\bar{\gamma} \bar{\partial}(-\frac{1}{\pi(x-\gamma)})
-\frac{1}{2}
{\partial}_x^2\bar{\partial}(-\frac{1}{\pi(x-\gamma)})\partial \gamma \nonumber \\
& & \qquad \qquad \qquad \qquad +\partial_x\bar{\partial} (-\frac{1}{\pi(x-\gamma)})
\partial\phi +O(e^{-2 \phi})
\Big]\nonumber\\
&=& \frac{l^2}{\alpha'}\int d^2z \Big[ e^{2 \phi}  \partial\bar{\gamma} \bar{\partial}(-\frac{1}{\pi(x-\gamma)})
-\bar{\partial}(-\frac{1}{\pi(x-\gamma)^3})\partial \gamma -\bar{\partial} (-\frac{1}{\pi(x-\gamma)^2})
\partial\phi +\dots
\Big]
\, . \nonumber 
\end{eqnarray}
Finally, by partial integration and using the equations of motion once more we conclude
\begin{eqnarray}
{\cal T}_{xx}(x,\bar{x}) &=& \frac{l^2}{\alpha'}\oint\frac{dz}{2\pi i}
\Big[\frac{e^{2 \phi}  \partial\bar{\gamma}}{x-\gamma}
-\frac{\partial\gamma}{(x-\gamma)^3}
-\frac{\partial\phi}{(x-\gamma)^2}
\Big]+O(e^{-2 \phi})\text{ terms}\, .
\end{eqnarray}
This agrees with the boundary $\phi\to\infty$ limit of equation (6.2) of \cite{Kutasov:1999xu} (using the limiting behavior of their function $\Lambda \to \frac{1}{x-\gamma}$ and their equation (2.12)). Thus, the case of our analysis where we restrict to a planar boundary matches the literature.\footnote{The value of the curvature radius $l$ is $l=\sqrt{k \alpha'}$ in the $AdS_3$ background in \cite{Kutasov:1999xu}.} 

\section{The Conservation of the Energy-Momentum}
\label{Conservation}
In this Appendix, our goal is to prove the conservation of the energy-momentum tensor in a conformally flat background.
The idea of the proof is that the variation of the action with respect to a space-time diffeomorphism is proportional to the world sheet equations of motion. This is because a space-time diffeomorphism is a world sheet field variation  $\delta X^\mu = \xi^\mu$.
In an $AdS$ background, and given the relation between the space-time action and the world sheet partition function, we therefore expect  equality between the variation of the space-time action under boundary diffeomorphisms and the variation of the world sheet action under field variations: 
 \begin{equation}
\int_\partial d^2 x' \sqrt{g^{(0)}}(x') \nabla^{i \, {(0)}} {\cal T}_{ij} (x') \xi^j (x') 
= \int d^2 z \, \xi^{i} (EOM)_i \, . \label{MatchEOM}
\end{equation}
We shall compute the left hand side based on the expressions we found for the energy-momentum tensor components, and the right hand side directly using the world sheet action. The proof will consistent in the fact that the expressions agree and that therefore the operator $\nabla^{i (0)} {\cal T}_{ij} $ equals zero and the energy-momentum tensor is conserved.

We slightly deviate from the calculation in the bulk of the paper. We choose the conformally flat background, and consider the anti-symmetric two-form in the gauge:\footnote{ The gauge we use here is a generalization of the one used for the $AdS_3$ background in Appendix \ref{Reduction}. The difference with the bulk parameterization is an anti-symmetric gauge transformation which is  BRST trivial. Recall also from Appendix \ref{Reduction} that we defined a   coordinate  $e^{\phi}=r$.}
\begin{eqnarray}
B_{x\bar{x}} &=& \frac{1}{2}e^{2\phi +2 \omega} +O(e^{-2\phi} ) \nonumber\\
B_{\phi x} &=& 2\partial_x\omega \, .
\end{eqnarray}
We work to zeroth order in the radial coordinate.
It is sufficient to perform a boundary diffeomorphism with as only variation $\delta \bar{x} = \xi^{\bar x}$. 
When we differentiate (\ref{MatchEOM}) with respect to $\xi^{\bar x}$ on the left and right hand side, we find:
\begin{eqnarray}
\sqrt{g^{(0)}}(x') \nabla^{i \, {(0)}} T_{i \bar{x}} (x') &=& 
 \int d^2 z \delta(x-x') (EOM)_{\bar x} \, ,
\end{eqnarray}
where $(EOM)_{\bar x}$ is proportional to the equation of motion obtained by varying $\bar{x}$. A calculation in which we neglect subleading terms gives rise to:
\begin{equation}
 (EOM)_{\bar x} =
 \frac{l^2}{\alpha'}
\int d^2 z \delta(x-x') \Big( 
  \partial (e^{2 \phi+2\omega} \bar{\partial} x)
+ (\partial_x \partial_{\bar{x}}^2 \omega - 2 \partial_x \partial_{\bar x} \omega \partial_{\bar x} \omega ) \bar{\partial} \bar{ x}
\partial x
- 2 \partial_{\bar x} \partial_x \omega \bar{\partial} \phi \partial x
\Big) \,  .
 \label{VariationAction}
\end{equation}
The second step in our proof is to compute the operator that equals the covariant derivative of the energy-momentum tensor in the conformally flat background, and to show that it agrees with the variation of the action (\ref{VariationAction}). Thus, it will be exhibited to be a trivial operator, proving conservation of energy-momentum. 

The energy-momentum conservation operator equals:
\begin{eqnarray}
\nabla^i {\cal T}_{ij} (x', \bar{x}') =
\partial_{x'} T_{\bar{x} \bar{x}}(x',\bar{x}') + \partial_{\bar{x}'} T_{x \bar{x}}
- 2 \partial_{\bar{x}'} \omega T_{x \bar{x}} \, . \label{CovariantDerivative}
\end{eqnarray}
We proceed as in the bulk of the paper, but in a mildly different gauge.
The three form $H_{(3)}$ is given by 
\begin{eqnarray}
H &=& (2e^{2\phi}\sqrt{g^{(0)}}-\frac{1}{4}e^{2\omega}R^{(0)}+O(e^{-2\phi}))d\phi\wedge dx\wedge d\bar{x}
\end{eqnarray}
Exploiting the expression \eqref{RicciScalarPerturbation} for the Ricci scalar, we can pick the anti-symmetric tensor:
\begin{eqnarray}
B_{x\bar{x}} &=& e^{2\phi}\sqrt{g^{(0)}}+O(e^{-2\phi})\nonumber\\
B_{\phi x} &=& 2\partial_x\omega-e^{-2\omega}\partial_{\bar{x}}h_{xx}.
\end{eqnarray}
The equations of motion imply the  order estimates
\begin{equation}
\bar{\partial}x = O(e^{-2\phi})\, , \qquad
\partial\bar{x} = O(e^{-2\phi})\, , \qquad
\partial\bar{\partial}(\phi+\omega) = O(e^{-2\phi}) \, .
\end{equation}
Therefore, the leading (non-vanishing) order of the energy-momentum tensor is
\begin{eqnarray}
{\cal T}_{\bar{x}\bar{x}}(x') &=& \frac{l^2e^{2\omega(x')}}{\alpha'}
\int d^2z \Big[e^{2\phi}\delta^{(2)}(x'-x)\partial x\bar{\partial}x-\frac{1}{2}\partial_{\bar{x}}(e^{-2\omega(x)}\partial_{\bar{x}}\delta^{(2)}(x'-x))\partial x\bar{\partial}\bar{x}\nonumber\\
&& +e^{-2\omega(x)}(\partial^2_{\bar{x}}\omega-(\partial_{\bar{x}}\omega)^2)\delta^{(2)}(x'-x)\partial x\bar{\partial}\bar{x}-e^{-2\omega(x)}\partial_{\bar{x}}\delta^{(2)}(x'-x)\partial x\bar{\partial}\phi\Big] \, .
\label{InitialEnergyMomentumTensorComponent}
\end{eqnarray}
One can distribute the factor $e^{2\omega(x')}$ on each term and use that $e^{2\omega(x')}\delta^{(2)}(x'-x)=e^{2\omega(x)}\delta^{(2)}(x'-x)$ to  simplify this energy momentum component:
\begin{eqnarray}
{\cal T}_{\bar{x}\bar{x}}(x') &=& \frac{l^2}{\alpha'}
\int d^2z \Big[e^{2\phi+2\omega(x)}\delta^{(2)}(x'-x)\partial x\bar{\partial}x-\frac{1}{2}\partial^2_{\bar{x}}(\delta^{(2)}(x'-x))\partial x\bar{\partial}\bar{x} \label{Txbarxbar1} \\
&& -\partial_{\bar{x}}(\delta^{(2)}(x'-x))\partial x\bar{\partial}(\phi+\omega)-(\partial_{\bar{x}}\omega)^2 \delta^{(2)}(x'-x)\partial x\bar{\partial}\bar{x} -2\partial_{\bar{x}}\omega \delta^{(2)}(x'-x)\partial x\bar{\partial}\phi\Big] . \nonumber
\end{eqnarray}
Using the properties of the delta function, one can write
\begin{eqnarray}
{\cal T}_{\bar{x}\bar{x}}(x') &=& \frac{-l^2}{\pi\alpha'}
\int d^2z \Big[\partial(\frac{1}{\bar{x'}-\bar{x}})e^{2\phi+2\omega(x)}\bar{\partial}x-\frac{1}{2}\partial^2_{\bar{x'}}\partial(\frac{1}{\bar{x'}-\bar{x}})\bar{\partial}\bar{x} \label{Txbarxbar} \\
&& +\partial_{\bar{x'}}\partial(\frac{1}{\bar{x'}-\bar{x}})\bar{\partial}(\phi+\omega)+(\partial_{\bar{x'}}\omega)^2 \partial(\frac{1}{\bar{x'}-\bar{x}})\bar{\partial}\bar{x}-2\partial_{\bar{x'}}\omega \partial(\frac{1}{\bar{x'}-\bar{x}})\bar{\partial}(\phi+\omega)\Big] \, . \nonumber
\end{eqnarray}
Then, the $x'$ derivative of the energy momentum tensor which appears in the covariant derivative (\ref{CovariantDerivative}) is
\begin{eqnarray}
\partial_{x'}{\cal T}_{\bar{x}\bar{x}}(x') &=& \frac{-l^2}{\pi\alpha'}
\int d^2z \Big[\partial(\partial_{x'}\frac{1}{\bar{x'}-\bar{x}})e^{2\phi+2\omega(x)}\bar{\partial}x-\frac{1}{2}\partial(\partial_{x'}\partial^2_{\bar{x'}}\frac{1}{\bar{x'}-\bar{x}})\bar{\partial}\bar{x} \label{Conservation1}\\
&& +\partial(\partial_{x'}\partial_{\bar{x'}}\frac{1}{\bar{x'}-\bar{x}})\bar{\partial}(\phi+\omega)+ \partial\partial_{x'}(\frac{(\partial_{\bar{x'}}\omega)^2}{\bar{x'}-\bar{x}})\bar{\partial}\bar{x}
 -2 \partial\partial_{x'}(\frac{\partial_{\bar{x'}}\omega}{\bar{x'}-\bar{x}})\bar{\partial}(\phi+\omega)\Big] \, .
\nonumber
\end{eqnarray}
The second and third terms are trivial, because they are akin to pure gauge vector fields with well-defined gauge parameters, proportional to $\delta$ functions.\footnote{See e.g. \cite{Kutasov:1999xu} for an elaborate discussion of this point.} Of the remaining terms, we also consider only those that are not proportional to $\delta$ functions. We further manipulate:
\begin{eqnarray}
\partial_{x'}{\cal T}_{\bar{x}\bar{x}}(x') &=& \frac{-l^2}{\pi\alpha'}
\int d^2z \Big[\partial(\partial_{x'}\frac{1}{\bar{x'}-\bar{x}})e^{2\phi+2\omega(x)}\bar{\partial}x +\partial(\frac{\partial_{x'} (\partial_{\bar{x'}}\omega)^2}{\bar{x'}-\bar{x}})\bar{\partial}\bar{x} -2 \partial(\frac{\partial_{x'}\partial_{\bar{x'}}\omega}{\bar{x'}-\bar{x}})\bar{\partial}(\phi+\omega)\Big]\nonumber\\
&=& \frac{-l^2}{\pi\alpha'}
\int d^2z \Big[\partial(\partial_{x'}\frac{1}{\bar{x'}-\bar{x}})e^{2\phi+2\omega(x)}\bar{\partial}x+2\partial_{x'}\partial_{\bar{x'}}\omega \partial(\frac{1 }{\bar{x'}-\bar{x}})(\partial_{\bar{x'}}\omega)\bar{\partial}\bar{x} \nonumber\\
&& \qquad\qquad - 2\partial_{x'}\partial_{\bar{x'}}\omega\partial(\frac{1}{\bar{x'}-\bar{x}})\bar{\partial}(\phi+\omega)\Big]\nonumber\\
&=& \frac{-l^2}{\alpha'}
\int d^2z \partial(\delta^{(2)}(x'-x))e^{2\phi+2\omega(x)}\bar{\partial}x \nonumber\\
&&+\frac{2l^2}{\alpha'}\partial_{x'}\partial_{\bar{x'}}\omega\int d^2z \Big[\delta^{(2)}(x'-x) \partial_{\bar{x'}}\omega\partial x \bar{\partial}\bar{x} - \delta^{(2)}(x'-x)\partial x \bar{\partial}(\phi+\omega)\Big]\nonumber\\
&=& \frac{l^2}{\alpha'}\int d^2z \Big[\delta^{(2)}(x'-x)(\partial(e^{2\phi+2\omega(x)}\bar{\partial}x)-2\partial_{x}\partial_{\bar{x}}\omega\partial x \bar{\partial}\phi)
\nonumber\\ && -
\partial(\delta^{(2)}(x'-x)e^{2\phi+2\omega(x)}\bar{\partial}x)\Big] \, . 
\label{ordinaryderivativeofTxbarxbar}
\end{eqnarray}
 The last term in the last line is a well-defined total derivative term that we can neglect.
Next, we covariantly differentiate the component ${\cal T}_{x \bar{x}}$ and find the contribution:
\begin{equation}
\partial_{\bar{x'}} {\cal T}_{x \bar{x}} 
-2 \partial_{\bar x'} \omega {\cal T}_{x\bar{x}} =\frac{l^2}{\alpha'}\partial_{x'}\partial_{\bar{x'}}\partial_{\bar{x'}}\omega\int d^2z \delta^{(2)}(x'-x)\partial x
\bar{\partial}\bar{x} 
 -2 \partial_{\bar x'} \omega
\frac{l^2}{\alpha'}\partial_{x'}\partial_{\bar{x'}}\omega\int d^2z \delta^{(2)}(x'-x)\partial x \bar{\partial}\bar{x}
\label{CovariantDerivativeTrace}
\end{equation}
to the conservation equation.
Combining the terms (\ref{ordinaryderivativeofTxbarxbar}) and (\ref{CovariantDerivativeTrace}), we find that the covariant derivative (\ref{CovariantDerivative}) is equal to
\begin{equation}
\nabla^i {\cal T}_{ij}   
= 
\frac{l^2}{\alpha'}\int d^2z \delta^{(2)}(x'-x)
\Big[\partial(e^{2\phi+2\omega(x)}\bar{\partial}x)+\partial_x\partial_{\bar{x}}^2\omega\partial x \bar{\partial}\bar{x}-2\partial_x\partial_{\bar{x}}\omega
\partial_{\bar{x}} \omega\partial x \bar{\partial} \bar{x}
-2\partial_x\partial_{\bar{x}}\omega\partial x\bar{\partial}\phi\Big]  \, .  \nonumber
\end{equation}
This matches  the variation (\ref{VariationAction}) of the action  under an infinitesimal field redefinition. Thus we have shown that the energy-momentum tensor is conserved.

\section{The Twisted Trace}
\label{TwistedTrace}
The flat gauge fields that satisfy the boundary conditions (\ref{TwistedBoundaryAsymptotics}) for conformally flat metric (for a given choice of twist) are locally:
\begin{equation}
A^R = \frac{i}{2} d\omega \, , \qquad\quad
\bar{A}^R = \frac{i}{2} d\omega.
\end{equation}
The action becomes 
\begin{eqnarray}
S &=& \frac{1}{2\pi\alpha'}\int d^2z \Big[(G_{\mu\nu}+B_{\mu\nu})\bar{\partial}X^{\mu}\partial X^{\nu}+il\bar{\partial}\omega\partial\theta + il\partial\omega\bar{\partial}\theta+\partial\theta\bar{\partial}\theta\Big] \, . \label{TwistedAction}
\end{eqnarray}
The equation of motion for the angle  $\theta$ is
\begin{equation}
    \partial\bar{\partial}(\theta+ il\omega) = 0.
    \label{ThetaTwisted}
\end{equation}
The variation of the conformal factor in the boundary condition gives rise to an extra term in the energy-momentum tensor component $T_{x\bar{x}}^{top}$:
\begin{eqnarray}
T_{x\bar{x}}^{top} &=& T_{x\bar{x}}^{phys} + \frac{il}{2\alpha'}\int d^2z \Big[\bar{\partial}(\delta^{(2)}(x'-x))\partial\theta + \partial(\delta^{(2)}(x'-x))\bar{\partial}\theta\Big].
\end{eqnarray}
Using the equation of motion (\ref{ThetaTwisted}), and up to BRST exact terms, we can write
\begin{eqnarray}
T_{x\bar{x}}^{top} &=& T_{x\bar{x}}^{phys} + \frac{l^2}{2\alpha'}\int d^2z \Big[\bar{\partial}(\delta^{(2)}(x'-x))\partial\omega+\partial(\delta^{(2)}(x'-x))\bar{\partial}\omega\Big]\nonumber\\
&=& T_{x\bar{x}}^{phys} - \frac{l^2}{\alpha'}\int d^2z \delta^{(2)}(x'-x)\partial\bar{\partial}\omega\nonumber\\
&=& 0 \,  ,
\end{eqnarray}
up to subleading terms, of order $O(e^{-2 \phi})=O(r^{-2})$. Finally, let us remark that the twisted action (\ref{TwistedAction}) shows that the left-moving and right-moving part of the circular direction $\theta$ is shifted by $\omega$ upon twisting. When we perform the T-dual topological twist, it is convenient to perform the above reasoning using the T-dual coordinate $\theta_{dual}=\theta_L-\theta_R$. The underlying reason is that it is the T-dual circle that renders the corresponding (axial) $U(1)_R$ symmetry geometrical.

\bibliographystyle{JHEP}

\end{document}